\documentstyle[prd,preprint,tighten,aps,eqsecnum,floats,%
amssymb,amsfonts,newlfont,graphicx]{revtex}
\begin{document}
\draft
\preprint{
\begin{tabular}{r}
   DFTT 71/99
\\ hep-ph/9912427
\end{tabular}
}
\title{Neutrino oscillations and neutrinoless double-$\beta$ decay}
\author{C. Giunti}
\address{INFN, Sezione di Torino, and Dipartimento di Fisica Teorica,
Universit\`a di Torino,\\
Via P. Giuria 1, I--10125 Torino, Italy}
\maketitle
\begin{abstract}
We consider the scheme
with mixing of three neutrinos and a mass hierarchy.
We shown that,
under the natural assumptions that massive neutrinos are Majorana particles
and there are no unlikely fine-tuned cancellations
among the contributions of the different neutrino masses,
the results of solar neutrino experiments
imply a lower bound
for the effective Majorana mass in neutrinoless double-$\beta$ decay.
We also discuss briefly
neutrinoless double-$\beta$ decay
in schemes with mixing of four neutrinos.
We show that one of them is favored by the data.
\end{abstract}

\pacs{Presented at TAUP'99,
6--10 September 1999, College de France, Paris, France.}

Neutrino oscillations
\cite{BGG-review-98}
have been observed in solar and atmospheric neutrino experiments.
The corresponding neutrino mass-squared differences are
\begin{equation}
\Delta{m}^2_{\mathrm{sun}} \sim 10^{-6} - 10^{-4} \, \mathrm{eV}^2
\qquad
\mbox{(MSW)}
\,,
\label{dm2-sun-MSW}
\end{equation}
in the case of MSW transitions,
or
\begin{equation}
\Delta{m}^2_{\mathrm{sun}} \sim 10^{-11} - 10^{-10} \, \mathrm{eV}^2
\qquad
\mbox{(VO)}
\,,
\label{dm2-sun-VO}
\end{equation}
in the case of vacuum oscillations,
and
\begin{equation}
\Delta{m}^2_{\mathrm{atm}} \sim 10^{-3} - 10^{-2} \, \mathrm{eV}^2
\,.
\label{dm2-atm}
\end{equation}
These values of the neutrino mass-squared differences
and the mixing required for the
observed solar and atmospheric oscillations
are compatible with the simplest and most natural
scheme with three-neutrino mixing and a mass hierarchy:
\begin{equation}
\underbrace{
\overbrace{m_1 \ll m_2}^{\Delta{m}^2_{\mathrm{sun}}}
\ll m_3
}_{\Delta{m}^2_{\mathrm{atm}}}
\,.
\label{mass-hierarchy}
\end{equation}
This scheme is predicted by the see-saw mechanism \cite{BGG-review-98},
which
predicts also that the three light massive neutrinos are
Majorana particles.
In this case
neutrinoless double-$\beta$ decay ($\beta\beta_{0\nu}$)
is possible and its matrix element
is proportional to the effective Majorana mass
\begin{equation}
|\langle{m}\rangle|
=
\left|
\sum_{k}
U_{ek}^2
\,
m_{k}
\right|
\,,
\label{effective}
\end{equation}
where $U$ is the neutrino mixing matrix
and the sum is over the contributions of all the mass eigenstate
neutrinos $\nu_k$ ($k=1,2,3$).

In principle the effective Majorana mass (\ref{effective})
can be vanishingly small because of cancellations among the
contributions of the different mass eigenstates.
However,
since the neutrino masses and the elements of the neutrino mixing matrix
are independent quantities,
if there is a hierarchy of neutrino masses
such a cancellation would be the result of an unlikely fine-tuning,
unless some unknown symmetry is at work.
Here we consider the possibility that no such symmetry exist and
\emph{no unlikely fine-tuning operates
to suppress the effective Majorana mass} (\ref{effective})
\cite{Giunti-lbb-99}.
In this case we have
\begin{equation}
|\langle{m}\rangle|
\simeq
\max_k |\langle{m}\rangle|_k
\,,
\label{max}
\end{equation}
where
$|\langle{m}\rangle|_k$
is the absolute value of the contribution of the massive neutrino
$\nu_k$ to $|\langle{m}\rangle|$:
\begin{equation}
|\langle{m}\rangle|_k
\equiv
|U_{ek}|^2 m_{k}
\,.
\label{mk}
\end{equation}
In the following we will estimate the value of
$|\langle{m}\rangle|$
using the largest
$|\langle{m}\rangle|_k$
obtained from the results of neutrino oscillation experiments.

\begin{figure}[t]
\begin{center}
\mbox{\includegraphics[bb=95 520 323 737,width=0.95\linewidth]{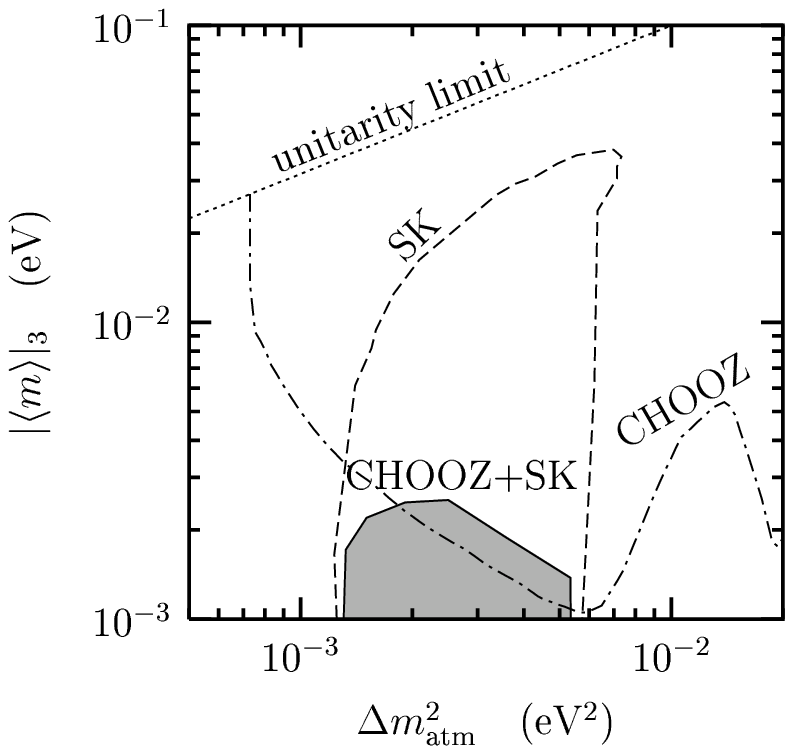}}
\\
\Large
Figure \ref{m3}
\end{center}
\refstepcounter{figure}
\label{m3}
\end{figure}

Let us consider first
$|\langle{m}\rangle|_3$,
which,
taking into account that in the three-neutrino scheme
under consideration
$m_3 \simeq \sqrt{\Delta{m}^2_{31}} = \sqrt{\Delta{m}^2_{\mathrm{atm}}}$,
is given by
\begin{equation}
|\langle{m}\rangle|_3
\simeq
|U_{e3}|^2
\sqrt{\Delta{m}^2_{\mathrm{atm}}}
\,.
\label{m3-def}
\end{equation}
Since the results of the CHOOZ experiment \cite{CHOOZ}
and the Super-Kamiokande atmospheric neutrino data \cite{SK}
imply that $|U_{e3}|^2$
is small
($|U_{e3}|^2 \lesssim 5 \times 10^{-2}$ \cite{Fogli}),
the contribution $|\langle{m}\rangle|_3$ to the effective
Majorana mass in $\beta\beta_{0\nu}$ decay
is very small
\cite{BGKM-bb-98,Giunti-lbb-99,BGGKP-bb-99}.
The upper bounds for
$|\langle{m}\rangle|_3$
as functions of
$\Delta{m}^2_{\mathrm{atm}}$
obtained from the present experimental data are shown in
Fig.~\ref{m3}.
The dash-dotted upper limit has been obtained using
the 90\% CL exclusion curve of the CHOOZ experiment
(taking into account \cite{BGG-review-98} that
$
|U_{e3}|^2
=
\frac{1}{2}
\left(
1 - \sqrt{ 1 - \sin^2 2 \vartheta_{\mathrm{CHOOZ}} }
\right)
$,
where $\vartheta_{\mathrm{CHOOZ}}$
is the two-neutrino mixing angle measured in the CHOOZ experiment),
the dashed upper bound
has been obtained using the results
presented in Ref.\cite{Fogli}
of the analysis
of Super-Kamiokande atmospheric neutrino data (at 90\% CL)
and the solid upper limit,
that surrounds the shadowed allowed region,
has been obtained using the results
presented in Ref.\cite{Fogli}
of the combined analysis
of the CHOOZ and Super-Kamiokande data (at 90\% CL).
The dotted line in Fig.~\ref{m3}
represents the unitarity limit
$
|\langle{m}\rangle|_3
\leq
\sqrt{\Delta{m}^2_{\mathrm{atm}}}
$.
One can see from Fig.~\ref{m3}
that the results of the CHOOZ experiment imply that
$
|\langle{m}\rangle|_3
\lesssim
2.7 \times 10^{-2} \, \mathrm{eV}
$,
the results of the Super-Kamiokande experiment imply that
$
|\langle{m}\rangle|_3
\lesssim
3.8 \times 10^{-2} \, \mathrm{eV}
$,
and the combination of the results of the two experiments
drastically lowers the upper bound to
\begin{equation}
|\langle{m}\rangle|_3
\lesssim
2.5 \times 10^{-3} \, \mathrm{eV}
\,.
\label{m3-max}
\end{equation}
Since there is no lower bound for $|U_{e3}|^2$
from experimental data,
$|\langle{m}\rangle|_3$
could be much smaller than the upper bound in Eq.~(\ref{m3-max}).

\begin{figure}[t]
\begin{center}
\mbox{\includegraphics[bb=95 520 323 737,width=0.95\linewidth]{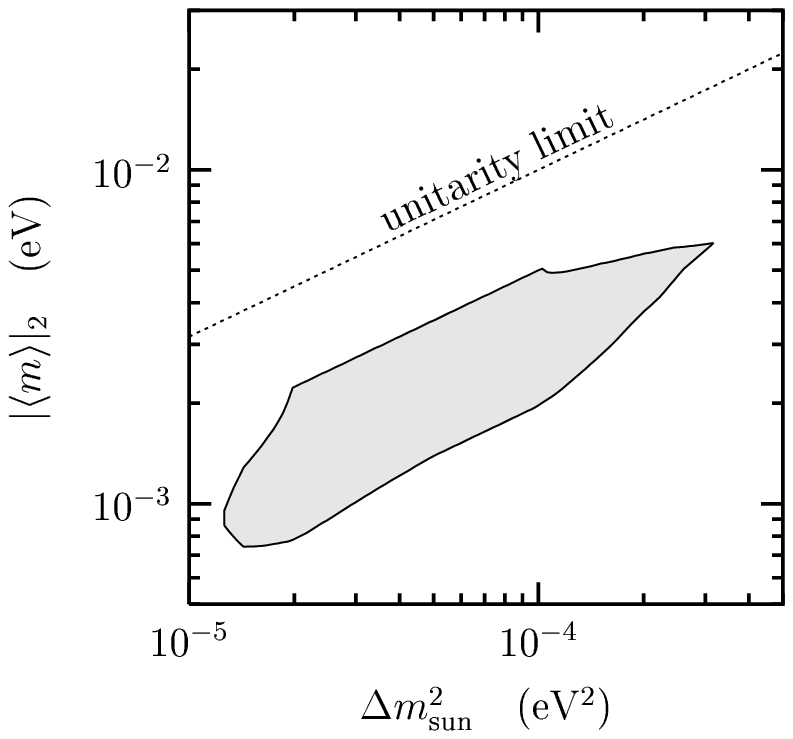}}
\\
\Large
Figure \ref{m2}
\end{center}
\refstepcounter{figure}
\label{m2}
\end{figure}

Hence,
the largest contribution to
$|\langle{m}\rangle|$
could come from
$
|\langle{m}\rangle|_2
\equiv
|U_{e2}|^2
\,
m_2
$.
In the scheme (\ref{mass-hierarchy})
$
m_2
\simeq
\sqrt{\Delta{m}^2_{21}}
=
\sqrt{\Delta{m}^2_{\mathrm{sun}}}
$
and,
since
$|U_{e3}|^2$
is very small,
$
|U_{e2}|^2
\simeq
\frac{1}{2}
\left(
1
-
\sqrt{ 1 - \sin^2 2\vartheta_{\mathrm{sun}} }
\right)
$
\cite{BG-98-dec},
where $\vartheta_{\mathrm{sun}}$
is the two-neutrino mixing angle used in the analysis of
solar neutrino data.
Therefore,
$|\langle{m}\rangle|_2$
is given by
\begin{equation}
|\langle{m}\rangle|_2
\simeq
\frac{1}{2}
\left(
1
-
\sqrt{ 1 - \sin^2 2\vartheta_{\mathrm{sun}} }
\right)
\sqrt{ \Delta{m}^2_{\mathrm{sun}} }
\,.
\label{m-2}
\end{equation}
Solar neutrino data imply bounds for $\sin^2 2\vartheta_{\mathrm{sun}}$
and
$ \Delta{m}^2_{\mathrm{sun}} $.
In particular the large mixing angle MSW solution (LMA)
of the solar neutrino problem
requires a relatively large
$\Delta{m}^2_{\mathrm{sun}}$
and a mixing angle $\vartheta_{\mathrm{sun}}$
close to maximal:
\begin{eqnarray}
&
1.2 \times 10^{-5} \, \mathrm{eV}^2
\lesssim
\Delta{m}^2_{\mathrm{sun}}
\lesssim
3.1 \times 10^{-4} \, \mathrm{eV}^2
\,,
\quad
&
\label{LMA-1}
\\
&
0.58
\lesssim
\sin^2 2\vartheta_{\mathrm{sun}}
\lesssim
1
\,,
\quad
&
\label{LMA-2}
\end{eqnarray}
at 99\% CL
\cite{SK-sun-analysis-99}.
The corresponding allowed range for
$|\langle{m}\rangle|_2$
as a function of
$\Delta{m}^2_{\mathrm{sun}}$
is shown in Fig.\ref{m2}
(the shadowed region limited by the solid line).
The dashed line in Fig.\ref{m2}
represents the unitarity limit
$ |\langle{m}\rangle|_2 \leq \sqrt{ \Delta{m}^2_{\mathrm{sun}} } $.
From Fig.\ref{m2}
one can see that the LMA solution
of the solar neutrino problem implies that
\begin{equation}
7.4 \times 10^{-4} \, \mathrm{eV}
\lesssim
|\langle{m}\rangle|_2
\lesssim
6.0 \times 10^{-3} \, \mathrm{eV}
\,.
\label{LMA-4}
\end{equation}
Assuming the absence of fine-tuned cancellations
among the contributions of the three neutrino masses
to the effective Majorana mass,
if $|U_{e3}|^2$ is very small and
$ |\langle{m}\rangle|_3 \ll |\langle{m}\rangle|_2$,
from Eqs.(\ref{max}) and (\ref{LMA-4}) we obtain
\begin{equation}
7 \times 10^{-4} \, \mathrm{eV}
\lesssim
|\langle{m}\rangle|
\lesssim
6 \times 10^{-3} \, \mathrm{eV}
\,.
\label{LMA-5}
\end{equation}
Hence,
assuming the absence of an
unlikely fine-tuned suppression of $|\langle{m}\rangle|$,
in the case of the LMA solution
of the solar neutrino problem
we have obtained a \emph{lower bound} of about
$7 \times 10^{-4} \, \mathrm{eV}$
for the effective Majorana mass
in $\beta\beta_{0\nu}$ decay.

Also the small mixing angle MSW (SMA)
and the vacuum oscillation (VO)
solutions
of the solar neutrino problem imply allowed ranges for
$|\langle{m}\rangle|_2$,
but their values are much smaller than in the case of the LMA solution.
Using the 99\% CL allowed regions
obtained in \cite{BKS-sun-analysis-98}
from the analysis of the total rates measured in solar neutrino experiments
we have
\begin{eqnarray}
5 \times 10^{-7} \, \mathrm{eV}
\lesssim
|\langle{m}\rangle|_2
\lesssim
10^{-5} \, \mathrm{eV}
& &
\mbox{(SMA)}
\,,
\label{SMA}
\\
10^{-6} \, \mathrm{eV}
\lesssim
|\langle{m}\rangle|_2
\lesssim
2 \times 10^{-5} \, \mathrm{eV}
& &
\mbox{(VO)}
\,.
\label{VO}
\end{eqnarray}
If future $\beta\beta_{0\nu}$
experiments will find
$|\langle{m}\rangle|$
in the range shown in Fig.\ref{m2}
and future long-baseline experiments
will obtain a stronger upper bound for
$|U_{e3}|^2$,
it would mean that
$|\langle{m}\rangle|_2$
gives the largest contribution to the effective Majorana mass,
favoring
the LMA solution of the solar neutrino problem.
On the other hand,
if future $\beta\beta_{0\nu}$
experiments will find
$|\langle{m}\rangle|$
in the range shown in Fig.\ref{m2}
and the SMA or VO solutions of the solar neutrino problem
will be proved to be correct by future solar neutrino experiments,
it would mean that
$|\langle{m}\rangle|_3$
gives the largest contribution to the effective Majorana mass
and there is a lower bound for the value
of $|U_{e3}|^2$.

Finally, let us consider briefly
the two four-neutrino mixing schemes compatible with all
neutrino oscillation data \cite{BGG-review-98},
including the indications in favor of $\nu_\mu\to\nu_e$
oscillations found in the short-baseline (SBL) LSND experiment
\cite{LSND}:
\begin{eqnarray}
\mbox{(A)}
&&
\underbrace{
\overbrace{m_1 < m_2}^{\Delta{m}^2_{\mathrm{atm}}}
<
\overbrace{m_3 < m_4}^{\Delta{m}^2_{\mathrm{sun}}}
}_{\Delta{m}^2_{\mathrm{SBL}}}
\,,
\label{A}
\\
\mbox{(B)}
&&
\underbrace{
\overbrace{m_1 < m_2}^{\Delta{m}^2_{\mathrm{sun}}}
<
\overbrace{m_3 < m_4}^{\Delta{m}^2_{\mathrm{atm}}}
}_{\Delta{m}^2_{\mathrm{SBL}}}
\,.
\label{B}
\end{eqnarray}
Since the mixing of $\nu_e$ with the two massive neutrinos
whose mass-squared difference generates atmospheric neutrino oscillations
is very small
\cite{BGG-review-98},
the contribution of the two ``heavy''
mass eigenstates $\nu_3$ and $\nu_4$
to the effective Majorana mass (\ref{effective})
is large in scheme A and very small in scheme B.
Hence,
the effective Majorana mass
is expected to be relatively large in scheme A
and strongly suppressed in scheme B.
In particular,
in the scheme A
the SMA solution of the solar neutrino problem
implies a value of $|\langle{m}\rangle|$
larger than the
the present upper bound obtained in
$\beta\beta_{0\nu}$ decay experiments
\cite{Baudis-99}
and is,
therefore, disfavored.
Furthermore,
since the measured abundances of primordial elements
produced in Big-Bang Nucleosynthesis
is compatible only with
the SMA solution of the solar neutrino problem \cite{BGGS-98-BBN},
we conclude that the scheme A is disfavored by the present experimental data
and
\emph{there is only one four-neutrino mixing scheme
supported by all data}:
scheme B
\cite{Giunti-lbb-99}.

\end{document}